\documentstyle[sprocl]{article}
\input epsf.sty
\def\Journal#1#2#3#4{{#1} {\bf #2}, #3 (#4)}

\def\PLB{{\em Phys. Lett.}  B}

\def\be{\begin{equation}}
\def\ee{\end{equation}}
\def\bea{\begin{eqnarray}}
\def\eea{\end{eqnarray}}

\begin{document}

\title{CONSTITUENT QUARKS FROM QCD:\\ 
                PERTURBATION THEORY AND THE 
                INFRA-RED~\footnote{Talk presented by 
E.~Bagan
}}

\author{ E.~BAGAN, M.~LAVELLE,  B.~FIOL, N.~ROY}

\address{Grup de F{\'{\i}}sica Te{\`o}rica, Departament de F{\'{\i}}sica and IFAE,\\ 
        Edifici Cn, Universitat Aut{\`o}noma de Barcelona \\
        E-08193 Bellaterra (Barcelona) Spain}

\author{D.~McMULLAN}

\address{Department of Mathematics and Statistics,\\ University of 
Plymouth, Drake Circus,\\ Plymouth, Devon PL4 8AA, UK}


\maketitle\abstracts{
Systematic approaches to building
up gauge invariant descriptions of charged fields, such
as electrons or quarks, are
described. Physically relevant descriptions must then be singled out
from a multiplicity of possibilities and to this end we give
a physical interpretation of one description. Perturbative
calculations which back up this interpretation are
outlined. A non-perturbative obstruction to observing an isolated
quark is reported. This sets the limits of the
constituent quark model.}
  

The physical content of our theories of the fundamental interactions
is profoundly affected by the gauge symmetries that lie at their heart.
Such theories are in
fact examples of systems with constraints and it is
well known that a consequence of this for QED is that only two of the
four initial $A^\mu$ potentials are actually physical. The implications
of the gauge symmetries of QED for charged fields (such as electrons)
are less well understood and for non-abelian theories, such as QCD, the
extraction of the physical degrees of freedom has not been performed.
This talk reports recent progress in understanding these
fundamental issues. In particular a gauge invariant
description of charged fields in electrodynamics and a
physical interpretation is provided. This leads to predictions which are
then tested in perturbation theory. The gauge structure of scalar QED
is so similar to that of standard QED that exactly
the same predictions may be made for it. They are also verified
here.

Any description of a physical charge must be gauge invariant and Gau\ss's law implies
an intimate link between charges and a chromo-(electro-)magnetic cloud. Such a 
description in terms of a so-called dressed field, $\psi_f$, was proposed by Dirac in the 50's: 
$
\psi_{f}(x)=\exp\left\{ ie\int d^4zf_\mu(z,x)A^\mu(z)\right\} 
\psi(x)
$. This can be seen to be gauge invariant if $f_\mu(z,x)$ satisfies
$\partial_\mu^z f^\mu(z,x)= 
\delta^{(4)}(z-x)$.

\begin{figure}
\centerline{\hbox{\epsfysize=5cm
\epsfbox{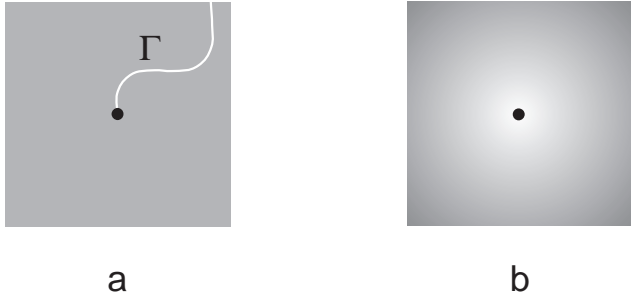}
}}
\caption{Two gauge invariant configurations of  a static field surrounded by a 
electromagnetic field. A stronger field is represented by a whiter shading. 
The very singular string-like configuration in (a) can be shown to decay into the
Coulombic one in (b) which is stable~\protect{\cite{www}}. 
\label{fig}}
\end{figure}

Not all gauge invariant descriptions are, however, 
physically relevant. In Fig.\ref{fig} two such
possible clouds are shown. 
Clearly, (a) is not stable in QED and for a static charge it will decay into the Coulomb
cloud in~(b). Our claim~\cite{LaMcMu96a,slow,fast} is that the latter, 
and a more general version corresponding to a charge moving with velocity
$\vec v$, are suitable for constructing physical asymptotic states. This in either QED or scalar 
QED since the magnetic field associated with the magnetic moment of an electron falls off rapidly 
away from the charge and is thus infra-red safe. 
To be explicit, for an electron moving with velocity $\vec v$ we 
dress the fermion as follows
\begin{equation}
\psi_v=\exp\left\{
ie{g^{\mu\nu}-(\eta+ v )^\mu(\eta- v)^\nu\over\partial^2-
(\eta\cdot\partial)^2+(
v\cdot\partial)^2}\partial_\nu A_\mu
\right\}
\psi ,
\label{eq:boos}
\end{equation}
where $v=(0,\vec v)$, and $\eta=(1,\vec 0)$.
Rather than giving the 
arguments underlying this statement, we now report a perturbative calculation~\cite{slow,fast} 
which verifies the following.   
Recall that the usual gauge-dependent fermion propagator in QED or QCD is plagued with infra-red 
divergences in an on-shell scheme. 
These reflect the fact that the fermion is not a good physical state ---the 
chromo-(electro-)magnetic field it generates is missing. If our dressing has a physical 
significance, we should be able to perform an {\em infra-red finite} 
on-shell renormalization for the propagator of the dressed charge defined by~(\ref{eq:boos}). 

For this non covariant description,  we find~\cite{fast} 
a multiplicative {\em matrix} renormalization
\begin{equation}
\psi^{({\rm bare})}_v=\sqrt{Z_2} \exp\left\{
-i { Z'\over Z_2}\sigma^{\mu\nu}\eta_\mu v_\nu \right\}\psi_v
\quad\mbox{\rm and}\quad m^{({\rm bare})}=m-
\delta m
\label{eq:Zs}
\end{equation}
necessary.
This is reminiscent of a naive Lorentz boost upon a fermion. The mass shift renormalization 
from demanding that the pole is at $m$ 
yields the standard gauge-invariant result found in any textbook.  The residue renormalization 
condition is where the infra-red divergences are usually found and we find here for our 
non covariant case three equations for only two unknowns, $Z'$ and $Z_2$. It is highly 
gratifying that at the  expected physical momentum, $p=m\gamma(1,\vec v)$, we can consistently 
solve these three equations and further that {\em the renormalization constants are gauge-invariant 
and infra-red finite}. 

The matrix renormalization~(\ref{eq:Zs}) is forced upon us by the fermion structure of QED. We 
have checked that in scalar QED where such a scheme is not possible, a straightforward 
multiplicative renormalization also yields infra-red finite results --- this 
despite exactly the same non-covariant dressing also being used in the scalar theory.  

The next step is the non-abelian theory. Quarks and gluons are believed to be confined inside
colourless hadrons and yet the success of the constituent quark model
and the jet structures observed in experiments show that it must be
possible to attach some physical meaning to quarks and gluons. The
Lagrangian fields are, however, gauge dependent ---like their QED
counterparts--- and we need to find the physical degrees of
freedom of such a non-abelian gauge theory. We note here that the action of the gauge dependent 
colour charge operator is only gauge invariant on locally gauge invariant objects and so the 
colour statistics of the constituent quark 
model require a gauge invariant description~\cite{LaMcMu96a}. 
Perturbative calculations are possible here, but, it can be shown that a 
non-perturbative obstruction, the Gribov ambiguity, prevents quarks and gluons from being true
observables. This is because the QCD equivalent of the dressing function $f_\mu(z,x)$  
can be used to fix the gauge. This obstruction 
sets the fundamental hadronic scale and the limits of the quark model.

\end{document}